\documentclass[review]{elsarticle}

\usepackage{t1enc}
\usepackage{amsmath}
\usepackage{amsfonts,dsfont}
\usepackage{amsthm}
\usepackage{amssymb}
\usepackage{graphicx}
\usepackage{multirow}
\usepackage{natbib}
\usepackage{lineno}
\usepackage{tikz,pgf}
\usetikzlibrary{arrows}
\usepackage{hyperref}
\hypersetup{colorlinks,
            linkcolor=blue,
            anchorcolor=blue,
            citecolor=blue,urlcolor={blue}}

\usepackage{booktabs}
\usepackage{siunitx}
\usepackage{threeparttable}
\newcommand{\sym}[1]{\ifmmode^{#1}\else\(^{#1}\)\fi} 

\modulolinenumbers[5]

\biboptions{authoryear}

\newtheorem{hypo}{Hypothesis}

\begin{document}
\begin{frontmatter}
\title{Impact of AI Tools on Learning Outcomes: Decreasing Knowledge and Over-Reliance}
\tnotetext[mytitlenote]{Preliminary version, DO NOT CIRCULATE.}

\author[corv,gt]{M\'arton Benedek}
\author[gt,corv]{Bal\'azs R. Sziklai}

\newtheorem{theorem}{Theorem}
\newtheorem{lemma}[theorem]{Lemma}

\address[corv]{Corvinus University of Budapest, Department of Operations Research and Actuarial Sciences, H-1093 Budapest F\H{o}v\'am t\'er 8.}
\address[gt]{Institute of Economics, ELTE Centre for Economic and Regional Studies, Email: sziklai.balazs@krtk.elte.hu}

\begin{abstract}
Students at all levels of education are increasingly relying on generative artificial intelligence (AI) tools to complete assignments and achieve higher exam scores. However, it remains unclear how this reliance affects their motivation, their genuine understanding of the material, and the extent to which it substitutes for the process of knowledge acquisition.
To investigate the impact of generative AI on learning outcomes, an experiment was conducted at Corvinus University of Budapest. In an operations research class, students were randomly assigned into two groups: one was permitted to use AI tools during classes and examinations, while the other was not. To ensure fairness, a compensation mechanism was introduced: students in the lower-performing group received point adjustments until the average performance of the two groups was equalized. 
Despite the organizers’ best efforts to explain the design and to create equal opportunities for all participants, many students perceived the experiment as a major disruption. 
Although the experiment was approved by every relevant university authority---including the Ethics Board, the Head of Department, the Program Director, and the Student Council---students escalated their concerns to the media and eventually to the State Secretary for Higher Education of Hungary.
As a result, the experiment had to be substantially revised before completion: on the final exam the test group was merged with the control group. Still, the data allowed us to draw decisive conclusions regarding the students' learning habits. Uncontrolled use of AI tools leads to disengaged students and low understanding of material. The extreme reactions of the students proved even more revealing than the data collected: generative AI tools have already become indispensable for students, raising fundamental questions about the validity of their learning process.
\end{abstract}

\begin{keyword}
artificial intelligence; learning outcomes; randomized experiment  \\
\end{keyword}


\end{frontmatter}


\normalsize

\section{Introduction}



Many important challenges that young people face are related to education. Generative AI applications excel at writing essays and are useful at various assignments including programming tasks. Unsurprisingly, students were among the first to adopt this new technology. A 2025 survey finds that 85\% of the students use AI in the U.S.\footnote{See \url{https://menlovc.com/perspective/2025-the-state-of-consumer-ai}.}

How the emergence of AI impacts students' learning processes is an important question that has attracted the interest of academics around the globe. The reports, however, are highly contradictory.
On the one hand, a recent meta-analysis of 51 experimental studies on the impact of ChatGPT found that AI has a large positive effect on improving learning performance and a moderately positive effect on enhancing learning perception \citep{Wang2025}. In a more recent experiment, \cite{Kestin2025} find that students learn significantly more in less time when using the AI tutor, compared with the in-class active learning.

On the other hand, emerging skeptical voices argue that such optimistic reports are at odds with established cognitive and behavioral models. The Law of Less Work (also known as the Principle of Least Effort) is a well-established observation in psychology stating that people tend to avoid options that involve more effort. Experimental evidence shows that this tendency extends to cognition as well: when given a choice, individuals systematically avoid cognitive demand \citep{kool2010}. Similarly, \cite{Westbrook2013} find that people act as 'cognitive misers', often willing to forgo substantial rewards in order to conserve mental effort.

Learning processes that are more effective are typically more costly as well, hence students shy away from them. Learners often do not appreciate the beneficial effects of the so-called desirably difficult tasks that are experienced as effortful, but have a positive effect on learning results and transfer of knowledge and skills \citep{Biwer2020,deBruin2023}. For example, retrieval practice is significantly more effective than simply re-reading material, however, self-testing requires greater cognitive effort \citep{Karpicke2009}.

These results prognosticate that AI tools, which produce results with minimal mental effort will be (i) popular among students and (ii) less effective in transforming the learning experience into lasting knowledge or skills. \cite{Gerlich2025} conducted surveys and in-depth interviews with 666 participants across diverse age groups and educational backgrounds and found a significant negative correlation between frequent AI tool usage and critical thinking abilities, mediated by increased cognitive offloading. In a randomized controlled experiment, \cite{Georgiou2025}  measured cognitive engagement of students during academic writing task assigning the participant into an AI-assisted and non-assisted (control) group. The results revealed significantly lower cognitive engagement scores in the AI-assisted group compared to the control group---or as the author paraphrases it: "ChatGPT produces more lazy thinkers". In a review paper, \cite{favero2025} warns that cognitive offloading undermines learning outcomes if the mental effort that is freed is not redirected towards other meaningful tasks. They also draw attention to further negative effects, such as superficial-learning, loss of agency, over-dependency, over-trusting and intellectual conformity.

\cite{Al-Zahrani2024} conducted a systematic review on the possible negatives effects that may accompany the integration of AI tools into education. The study uncovered concerns such as loss of human connection, algorithmic bias or reduced critical thinking and creativity among others. A subsequent survey with 260 participants from a Saudi Arabian university corroborated the findings and confirmed these concerns as valid.

Finally, \cite{kosmyna2025} conducted a laboratory experiment to examine the neural and behavioral consequences of LLM-assisted essay writing. Participants were divided into three groups: LLM, Search Engine, and Brain-only (no tools). Cognitive activity decreased when participants relied on external tools, and self-reported ownership of the essays was lowest in the LLM group and highest in the Brain-only group. Although LLMs provide immediate convenience, the findings underscore potential cognitive costs: over a four-month period, LLM users consistently underperformed across neural, linguistic, and behavioral measures.

In this paper, we report on a randomized experiment that aimed to measure the impact of AI tools on learning outcomes. Students of an operations research course in a Business Informatics program had been divided into two groups. The first group was allowed to use generative artificial intelligence tools (e.g. ChatGPT) both during class and in the exam, while the second group was not. Group assignment was carried out using a random algorithm to avoid self-selection bias. To ensure fair treatment, point compensation was applied between the groups: if one group performed worse, its students received the difference in points between the two groups’ average scores.

The experiment had to be revised due to student complaints, yet there are valuable lessons that can be derived both from the existing data and from the way the events unfolded. The paper can be divided into two parts. In Section~\ref{sec:Exp_main}, we provide a detailed account of the experiment. We present our hypothesis and the path that led to its formulation. How the experiment unfolded and its repercussions constitutes a case study in its own right. Then in Section~\ref{sec:Results}, we present the data salvaged from the revised setup. Although the treatment and control groups had to be merged, the analysis still offers a telling picture about the learning process of the students.

\subsection{Addressing the controversy in the literature}\label{sec:controversy}

We cannot present our results responsibly without addressing the controversy and ongoing academic debate surrounding this topic. The optimistic \citep{Rodrigues2025} and pessimistic \citep{favero2025} positions are so far apart that presenting evidence for only one side is insufficient. We aim to provide credible explanations for the observed false positives.

As described in Section~\ref{sec:Exp_design} the authors hypothesize that the truth is closer to the pessimistic view. There is a growing number of literature that strengthens the skeptical side. However, one resolution to the conundrum is that the skeptics are simply wrong. This can happen for multiple reason.

\begin{itemize}
    \item Generation Z (born roughly between 1997 and 2012) are generally more adept at learning with digital tools, and cognitive or behavioral patterns such as the Law of Less Work may not apply to them in the same way as to earlier generations.
    \item  Many studies advocate that the correct use of AI applications is essential for achieving positive learning outcomes. Perhaps, in experiments that reported positive effects on learning, instructors leveraged a way for student to meaningfully engage with AI, which induced active learning.
\end{itemize}

However, a number of reasons suggest that the optimism is unsubstantiated.

\begin{itemize}
    \item Improved performance might be due to sample bias. Top universities accommodate more motivated students who might use AI tools more responsibly.
    \item Experimenters might have misattributed performance improvement of students for various reasons. Detecting AI use is a difficult task, and academic dishonesty has never been more easy for students.
    \item Experiments with inadequate incentives (e.g.\ money reward) might result in false or misleading outcomes, while organizing experiments with natural incentives (grades) are challenging---as we will see in this paper.
    \item The authors may have conflated their own gains in productivity with those of the students. Generative AI tools can indeed serve as excellent tutors, particularly when learners are free from the pressure of tests and examinations, are driven by genuine curiosity, and possess the skills to validate the content provided by LLMs.
    \item Certain disciplines might be more resistant to misuse of AI.
    \item Finally, the new technology generated a considerable hype. In this environment, it is perhaps easier to publish results that support the positive aspects of the technology.
\end{itemize}

At this stage, we cannot tell which factors, individually or jointly, drive the discrepancies across experimental results, although we believe that the issue of incentives is the most serious, and most likely culprit. In the following, we describe how we set out to test the impact of generative AI tools.

\section{Experiment}\label{sec:Exp_main}

This section first provides valuable context for the experiment and sheds light on students’ general behavior toward AI. We then describe the experimental design and discuss how it unfolded as well its repercussions.

\subsection{Background and Personal experiences}

Operations research (OR) is an applied branch of mathematics that addresses topics such as optimal logistics, scheduling and production, using methodologies drawn from linear programming, graph theory, and related fields. With the rise of open-source programming languages and their accompanying software libraries, teaching practices in this discipline have evolved considerably. Today, modern operations research courses are often taught within programming environments. This shift comes with a few noteworthy implications.
Firstly, students face two levels of abstraction. One, when they transform an economic or business problem into a mathematical model, and another one when they implement their solution in a programming language. Not surprisingly, such a course is among the most challenging ones that students face during a typical business school program.

Secondly, the theory behind the mathematical models is relatively old and well-understood. Hence, even before the advent of generative AI tools, there were a large number of online educational materials that covered all elements of a typical OR course including programming exercises. As a consequence, generative AI models, which were trained on math related Q\&A sites such as StackExchange and StackOverflow, are generally well-versed in OR related questions. More precisely, there was a gradual improvement in the expertise and efficiency of generative AI models. By the time the experiment took place, the spring semester of 2025, the models could practically guarantee a passing grade.

The authors teach an OR course each spring in a Business Informatics program serving approximately 100-120 students at Corvinus University of Budapest (CUB). The course involves weekly seminars held in a computer lab and lectures on every second week. In the 2022/23 semester the course was redesigned to keep pace with advances in software technology. Seminars were taught using Jupyter Notebook environment of Python via the PuLP and NetworkX modules. To mitigate difficulties arising from double abstraction and to focus on developing students' modelling skills rather than rote recall, the authors administered open-book exams. Students were free to use any course materials or online resources during exams, including generative AI tools. The only restriction was that students had to answer all questions on their own; they were not permitted to receive help from any other \emph{person}.

During the 2022/2023 academic year very little out of the ordinary occurred. Innovative students experimented with AI tools, but they were a clear minority. At that time, AI models' capabilities were insufficient to substitute for learning the material.

In the 2023/2024 semester the scales have tipped in both areas. Significant number of students utilized some help from generative AI models, which in turn had improved enough to be an effective tool. A major incident---one that prompted the authors to design the experiment---occurred at the final exam. A student submitted an exam sheet that contained not only the solutions to his own personalized assignments but also those of other students alongside a considerable amount of gibberish. What initially appeared to be external manipulation turned out to be simple negligence on the student's part: instead of using the designated software to generate his own exercise, he fed the entire program (generating multiple exam sheets) to a generative AI agent and copied the output without reading any of it.

This was the moment the authors realized that students would not use AI models responsibly. As always when behavior is concerned, it is worth to examine the incentives. Students are primarily interested in securing appropriate grades---passing grades for less diligent students, and good or excellent grades for those motivated by scholarships or other academic recognition. Mathematics---even the applied kind---has, among many students, a reputation for lacking practical value. Consequently, many seek to minimize their effort while progressing through math courses, and generative AI models offer a powerful tool for achieving this goal.

In the 2024/25 semester---the term in which the experiment was conducted---adoption of generative AI among students in the course was virtually universal. Moreover, clear signs of over-reliance began to emerge. The major incident of the previous year became the new norm. Students happily substituted AI for the difficult work of learning, ceding control and allowing it to take a chance at securing a good grade with minimal or no effort. The experiment was designed precisely to capture this phenomenon.

\subsection{Experimental design}\label{sec:Exp_design}

Students in the Operations Research (OR) course were randomly assigned to treatment and control groups in a 1:1 ratio within each seminar groups. The allocation was stratified by gender (based on the name of students).
The treatment group was permitted to use generative AI tools during classes and exams, whereas the control group was not.\footnote{Labeling the AI-permitted group as the treatment is somewhat arbitrary, given that AI use was the norm in the previous year. Students were not informed which group was the treatment and which was the control and assumed that using AI is the natural way of things, thus they framed the non-permitted group as the treatment.} Since it is practically impossible to keep track of the wide variety of AI tools, the control group wrote each test in an offline computer lab. Both groups could, however, use lecture notes or any static offline material prepared beforehand. Henceforward, we will refer to the treatment and control as AI and offline groups respectively.

There were five seminar groups, three held by one of the authors and two by the other. Students were informed about the experiment on the first week, and assigned into groups in the second week. Participation was voluntary, but opting out only meant refusing data collection; group assignment remained compulsory, as it was integrated into the course requirements.

Students could earn up to 100 points during the semester. No extra credit assignment was planned to ensure that incentives are unaffected from external factors. The total score of the student consisted of the following components

\begin{itemize}
    \item First midterm on week 5 (18p),
    \item Second midterm on week 10 (18p),
    \item Short True or False paper test during final exam (10p),
    \item Final exam in the lab (54p).
\end{itemize}

The most important measurement was the paper test, which both groups took offline. It was designed to be of small importance---students could get top grades with scoring 0\% on the paper test and less than 100\% in all others. It contained true or false questions that tested only the basic understanding of the material. Our assumption was that by the Low of Less Work students in the AI-permitted group will ignore the test and rely completely on AI tools to obtain an appropriate grade. Students in the non-permitted group, however, need to engage with the material more deeply which will show on the test scores. This difference in knowledge-levels can manifest in two ways. First, we expect offline students to obtain better paper test scores than students in the AI group. Second, the obtained scores should align more closely with results of the final exam for the offline group, that is, more knowledgeable students are expected to perform better on other tests as well. Thus, we have formulated two hypotheses.

\begin{hypo}\label{hypo1}
There will be a statistically significant difference in paper test scores between the two groups. Specifically, students who were not permitted to use AI assistance (the offline group) will achieve significantly higher scores than those who were allowed to use AI tools.
\end{hypo}

To assess group differences, a permutation test was planned, given the lack of reliable information regarding the distribution of test scores.

\begin{hypo}\label{hypo2}
The alignment between paper test scores and final exam scores will be stronger for the offline group than for the AI-assisted group.
\end{hypo}

Alignment can be measured by how the test scores rank the students, hence statistical difference was assesed by the Sum of Ranking Differences procedure \citep{Heberger2010}. A linear regression analysis was performed to further validate the results.

The experiment was pre-registered at \cite{aspredicted223975}.

\subsection{Ethical considerations}\label{sec:ethics}

Since the experiment had to be substantially revised in response to student complaints shortly before the final data collection, it is worth reflecting on what went wrong. The authors emphasize that they strove to conduct a fair experiment and carefully considered its ethical aspects. The experiment was formally approved by

\begin{itemize}
    \item the Ethics Board of CUB,
    \item Vice Rector of Student Affairs at CUB,
    \item the Head of Institution of Operations and Decision Sciences of CUB,
    \item the Head of Department of Operations Research and Actuarial Sciences of CUB,
    \item the Program Director of the Business Informatics major of CUB, and
    \item the Student Council of CUB.
\end{itemize}

Furthermore, the planned experiment was presented in advance to the Centre for Quality Enhancement and Methodology in Education of CUB, to the dean responsible for Artificial Intelligence (CUB), and it was also discussed at several professional forums (e.g., the Budapest Experimental and Behavioral Economics Seminar at CUB, and the ELTE KRTK KTI mini-seminar series). The suggestions and insights raised there were incorporated into the experiment.

There are two major issues that were voiced by the students. The first is participation. Randomized group assignment was necessary to avoid self-selection bias and to ensure equal-sized treatment and control groups. What would mean opting out from the experiment in this context? If opting out means that students are prohibited using AI, then there is no real choice, participation is practically forced. If, instead, opting out meant being permitted to use AI, nobody would choose to participate. Ironically, a third option---allowing students who do not want to participate take an oral exam at the end of year---was discarded by the authors as dishonest: the course is taught primarily in a computer-lab environment, and few students would elect an oral assessment of such material. As we will see later, allowing this third option might have saved the experiment from a legal point of view.

To avoid this conundrum, the authors reorganized the course and made group assignment part of the examination process rather than an element of the experiment. While this may appear to compromise fairness, note that many popular, well-established assessment formats also undermine equity. For example, in group projects, students are randomly assigned to groups of 4–6; those paired with more prepared and diligent peers often earn better grades with less effort than those grouped with less prepared or less motivated peers. In exams with multiple questions sets (e.g.\ Forms A/B), one version may be easier despite the instructors' best efforts to balance difficulty. Finally, oral exams administered by multiple instructors risk inequity, as instructors differ in leniency and expertise. Indeed, the literature has long explored the impact of peer effects \citep{Booij2016}, methods of test equating to mitigate score differences \citep{kolen2014test}, and approaches to address rater effects arising from examiner stringency or leniency \citep{Harasym2008}.

In our case, we were well-aware that the introduction of AI creates an imbalance. To restore equity among students the authors introduced a point compensation scheme, that ensured that both groups have the same average performance. After each test, students in the lower performing group were compensated by the point difference between the group averages. Note that the compensation scheme was open for both groups. Indeed, there were instances where students in the AI-permitted group received compensation\footnote{Midterms featured different question sets, so groups were further divided by which test sheet they wrote. One complaint of the students was that compensation was based on too small groups. The authors addressed this by preparing fewer question sets for the second midterm.}.

Another important aspect is that compensation leaves incentives unaffected: each student is better off by studying. Although in theory, students in one group could cooperate and turn in empty sheets and receive the average score of the other group as compensation, this is unlikely as it is not a Nash-equilibrium point: any student could deviate from this strategy and the additional points he received by studying would be more than he lost on the diminishing compensation. In other words, each point earned in the test worth more than the possible compensation a student would get by not earning that point, and that was the authors intent precisely.

At this point, we invite the reader to ask: which---if any---group was treated unfairly? A natural first reaction is that students in the AI-permitted groups are better off, since they can always choose not to use AI. However, given our hypothesis---grounded in the Law of Less Work---that AI-permitted students engage less with the material and thus acquire less knowledge, while earning on average the same scores as the non-permitted group (owing to compensation in the assessment design), one can argue that the AI-permitted group is in fact the more disadvantaged. The whole point of the experiment was to uncover which group is ultimately better off in terms of knowledge.

Which brings us to the second issue. Students did not contest the unfairness of the compensation scheme or the grade, at least not at first. They did, however, point out that students in the AI-permitted groups need much less effort during studying, which translates into more free time. These resources can be used to get ahead in other subjects for example. Unfortunately, at the time of the experiment there was (and still is) an unhealthy competition for stipends. CUB students are eligible for stipends only if their credit-weighted average grade exceeds a specified threshold; the same applies to eligibility for tuition-free (similar to state-funded) status. Although the OR course carries only half the usual credit load (3 instead of 6), many perceived that AI-permitted students enjoyed an unfair advantage. The authors note that in the previous semester, when every student was allowed to use AI, grades still ranged from top marks to failures. Thus, access to AI did not eliminate the need for effort to earn high grades. Moreover, effort is an intrinsic, largely unobservable input and cannot be directly controlled or equalized. Unfortunately, these explanations did not manage to placate students.

\subsection{Unfolding events and Aftermath of the Experiment}

Despite the best efforts of the authors, the situation has slowly but steadily escalated until the experiment became a small national scandal\footnote{It even broke international news when Alvin Roth featured it in his seminal blog \url{https://marketdesigner.blogspot.com/2025/05/a-controversial-artificial-intelligence.html}}. On the second week, a few students complained in person about various aspects of the experiment. The authors managed to convince these individuals, however, the general dissent remained. Part of resentment was fueled by unfortunate group dynamics that was not revealed to the authors until too late. Allegedly, students in the AI-permitted groups started mocking non-permitted students in internal online forums.

Protests resurfaced after the first midterm. Students again questioned aspects of the experimental design. For example, a serious debate arose over why compensation was based on the mean rather than the median, which is more robust to outliers. The authors responded by facilitating open discussion and modifying parameters that were not critical to the design.

By the time of the second midterm a small but persistent fraction of the students started to appeal to university authorities, they approached the program director as well as the Dean responsible for Artificial Intelligence. As they did not succeed in overturning an approved experiment, confronted the instructors again. The heated discussions that followed were uncharacteristic to any experience the authors had with students before. Rational arguments were practically impossible as the students genuinely felt that they had been treated unfairly.

Finally, a group of students (approximately a dozen of them) turned to the media, and one of them even appealed to the State Secretary of Higher Education of Hungary. Telex, the largest online news portal of Hungary approached the University and interviewed the instructors as well as the students. The published report  
had an interesting effect. There was a moderate public uproar, one of the authors was even approached and bashed in social media for organizing of what seemed to be an unfair experiment. Fellow professors, and teachers in general, however, were perplexed by the students attitude. One student complaint, which was highlighted in the report read as follows.\hfill\\

"No amount of bonus points can make up for the extra hours someone spent studying, while the other group’s members finished in a few minutes with the help of AI." \citep{halasz2025_telex}\hfill\\

There is a lot to unpack in this remark. Ironically, the student complains about being forced to study. More importantly, however, the remark reveals that the students themselves believe the other group will not use AI responsibly, but instead as a substitute for learning the material.

Nevertheless, even before the media report erupted, the program director received the written opinion of the State Secretary, who expressed that the experiment even with the point compensation might harm the equal chances of the students and such harm is not justified by its goals. In light of this negative opinion the University decided to revise the experiment and merge the treatment and control groups. A telling nuance is the debate that emerged after this decision. Should the treatment group be collapsed into the control group, or \textit{vice versa}? In other words, after the merger, should all students be permitted to use AI, or should all students be prohibited? The authors argued that if restoring equity is the main concern, then prohibiting the use for all should be a better option. Students differ in economic background, and some of them can afford better AI tools (this is something the authors already knew at the planning of the experiment, but could not control for). However, the legal department of CUB discarded this idea, saying \textit{it is not equity that needs to be restored but rather the perception of equity}. 


As a result, one week before the exam period started, students were informed that the two groups (AI and offline) had been merged and that all students were permitted to use AI assistance during the exam. Moreover, they were given the option to discard their midterm test results and be graded solely on the basis of their final exam score. In practice, the instructors calculated the students' grades both ways and simply awarded the higher result.

Knowing that students prefer to concentrate the bulk of their studying efforts in the exam period, there was plenty of time \textit{not to prepare} for the exam. This makes it impossible to tell whether if there are no differences in the main measurements between the two groups it is because there never was, or because the experiment got compromised in the last week. It turns out, it does not really matter. The paper test scores were so abysmally low, that it is in itself a reliable proof of students' learning process, or rather the lack of it. Before, we turn to analyzing the data, let us draw some conclusions regarding the experiment.


\subsection{Lessons to be learned}\label{sec:exp_lessons}

Field experiments, especially behavioral ones are always difficult to administer. Experiments, even those designed in good faith with special attention to fairness aspects, might go astray in unexpected ways. Here we try to gather a few lessons that might prove useful to future experimenters.

\paragraph*{Perception}

In hindsight, the authors erred in focusing too much on real aspects of fairness, neglecting in how the experiment is perceived by the participants. Framing the experiment appropriately could have prevented the emergence of false narratives. The experiment also came as a surprise in the first week. Notifying students about a planned study before course registration might have reduced the shock value. The group dynamics and the existence of outside options are also related to perception, but due to their importance we treat them separately.

\paragraph*{Group dynamics}

In a field experiment, where the participants are monitored in a natural setting over an extended period of time, there are always some events that cannot be controlled for or even observed. Toxic group dynamics can seriously affect the outcome and even lead to extreme reactions. One way that the authors could have avoided such developments was to monitor the emotional conditions of the participants more closely (e.g. with weekly surveys, or discussion sessions).

\paragraph*{Outside option}

Even if the reader accepts the authors argument that encompassing group assignment into the course requirement was not problematic, providing an outside option would have been a much more prudent choice. The existence of an outside option also helps to defend the experiment from a legal point of view. The opinion of the State Secretary might have been less condemning and the perception of the experiment---both from the media and the student side---could have been better too.

\paragraph*{Design}

There are alternatives to the experimental design applied here that are less prone to criticism for creating unequal opportunities. One option is to reverse the roles of the students (AI vs.\ offline) midway through the semester, so that each student uses AI for half of the time. However, this design remains problematic if the difficulty of the material is uneven: students who are allowed to use AI during the more demanding period would be perceived as advantaged. Another possibility is to permit AI use during one part of the semester and prohibit it during the other for all students. While this approach treats all students equally, the results become harder to interpret, as they may simply reflect differences in the material rather than the impact of AI assistance.\\

The experiment raises questions about how one can ethically measure student behavior, as well as concerns about university autonomy. At what point should a university yield to media and government pressure? The authors and CUB are still processing the lessons.

Two immediate actions were taken to understand the events and to prevent their recurrence. First, the authors engaged an independent third party to run a focus group in which students could freely share experiences and grievances. Students could choose either a live moderator or an AI-based, trained chatbot \citep{Zrubka2025ModiBot}. Sadly, no one showed up on any of the sessions---which is curious given how strongly they felt about the issue, before their demand was accommodated. The second action involves the revision of the approval process for experiments conducted at CUB. The University established a body, independent of the Ethics Board, that will revise current procedures for evaluating experiment designs that involve sensitive elements.

\section{Results}\label{sec:Results}

During the experiment the following measurements were taken: first midterm test scores (AI and offline groups), second midterm test scores (AI and offline groups), paper test during final exam (offline for everyone), and final exam test scores (AI assistance permitted for everyone).

\begin{table}[t]
\centering
\begin{tabular}{l|ccccc|l}
                 & $n_1$ (AI) & $n_2$ (Offline) & AI    & Offline & Difference & p-value            \\ \hline
1st midterm      & 45         & 48              & 12.64 & 10.93   & 1.71       & 0.015*              \\
2nd midterm      & 44         & 48              & 11.51 & 5.93    & 5.58       & \textless 0.001*** \\
final exam       & 46         & 49              & 40.48 & 41.53   & -1.05      & 0.734      \\ \hline
true-or-false    & 46         & 49              & 30.08 & 27.71   & 2.37       & 0.011*
\end{tabular}
\caption{Permutation test results for the difference between group averages. Sample sizes ($n_1$, $n_2$) vary due to student attendance. Note that ‘true-or-false’ indicates the average number of correct answers that students achieved on the paper test, not the number of points they received for this task.}\label{tab:perm_test}
\end{table}

The primary measurement was the paper test, which was conducted as planned and administered offline for all students just before the final exam. However, the circumstances surrounding the test changed considerably. Students who had originally been assigned to the offline group were relieved to learn that they would not have to rely solely on their own knowledge. As a result, many did not feel the pressure to prepare for the exam, weakening our treatment effect and rendering our hypothesis very difficult to test. Not only did the offline group fail to outperform the AI-permitted group on the paper test, but there was in fact a slight, yet statistically significant, advantage in favor of the AI group (see Table~\ref{tab:perm_test}). Thus, Hypothesis~\ref{hypo1} was not supported by the data, although the causal relationship underlying the performance differences is clearly obscured by the merging of the treatment and control groups.

The AI group outperformed the offline group in the first and second midterm. From this data alone we should not draw any deep conclusions as we have no baseline to compare the two groups to. The difference might be simply due to the fact that generative AI tools are better at these kind of tasks than students. How the AI group would have fared without the LLMs assistance is unknown. The second midterm results suggest that the gap is widening, though this may also be explained by changes in the difficulty of the material, which could disproportionately affect offline students. Other factors related to the experimental setting---such as demotivation or deteriorating group dynamics---might also play a role. It is also notable, that the trend breaks on the final exam, where the offline group (armed with AI assistance) outperformed the AI group, although the difference is not statistically significant.

Still, there is something we can infer from the data. The experimental setting did not change for the AI group, so their paper test scores reflect their true knowledge level at the end of the semester. For the offline group, we expected performance to drop back to roughly the same level due to the merging of the two groups---and this indeed occurred, although the decline was even greater than anticipated, perhaps because the students felt overly relieved. In any case, the key question remains: how much do the students actually know?

\begin{figure}
    \centering
    \includegraphics[width=1\linewidth]{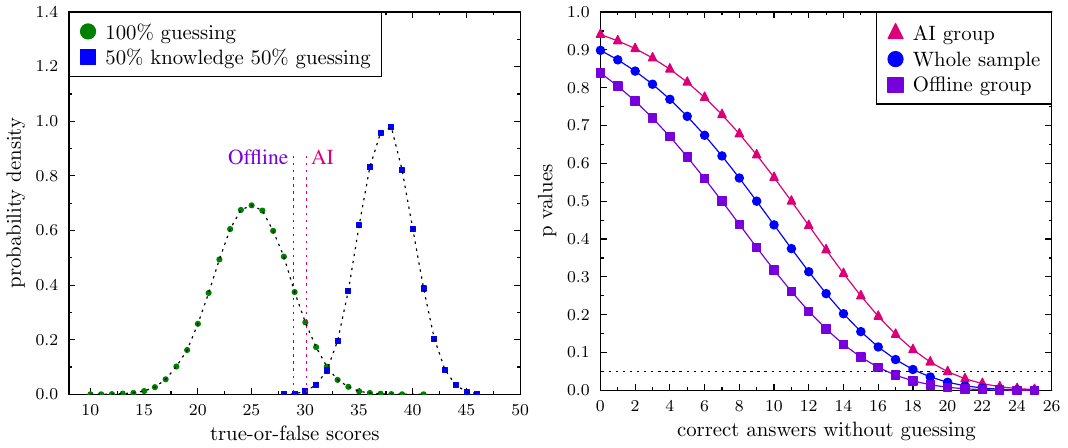}
    \caption{Students' grasp of the material as measured by the paper test. Neither group demonstrates knowledge levels significantly different from random guessing. Left: How paper test scores compare to knowledge levels modeled by binomial distribution. Right: Likelihood that, given the test scores, students knew $x$ questions correctly and guessed the remaining $50-x$.}
    \label{fig:paper_test}
\end{figure}


The paper test consisted of 50 true-or-false questions, grouped into ten topics, each containing five statements that the students had to be evaluated. For some of the topics the statements were only thematically related, for others the statements related to an exercise (for examples see the Appendix).

One way to model the students' grasp of the subject is to assume they have definitive knowledge about certain ratio of the material and for the rest they use random guessing. For simplicity, we assume that 'definitive' means 100\% correct answer for a statement, while random guessing happens to be true 50\% of the time. Hence, we can use the binomial distribution to predict the true-or-false test scores of the students. There are other ways to model test scores\footnote{Note, that random guessing is always available for the students, and evidence points toward that some of them indeed used this strategy. For instance, there were students who indicated both Statement b) and d) true in the paper test example 2 (see Appendix).}, the statements were not entirely independent and guessing might not happen with even chances, hence the distribution can be skewed or ragged, however, the significance thresholds are more or less robust to such changes.

Figure~\ref{fig:paper_test}, on the left, shows the average scores achieved on the paper test compared to binomial distributions corresponding to knowledge levels 0 and 50\%. On the right we can see the likelihood of obtaining such a test score ($P(X\le \overline{y})$ where $\overline{y}$ denotes the group average). If we draw the line at $\alpha=0.05$, students could not know more than 20 true-or-false questions correctly (40\% performance), but realistically their knowledge levels are closer to 20\% ($=$ ten correct answer) or lower. Although we cannot directly compare students' total scores with those of previous years (due to differences in exam format, student cohorts, and the presence of extra credit assignments), it is worth noting that no student could have passed without knowing at least 50\% of the material, and the average performance was well above that threshold. Thus, it is not unrealistic to claim that knowledge levels dropped by at least 20 percentage points, and likely more.

\begin{figure}
    \centering
    \includegraphics[width=1\linewidth]{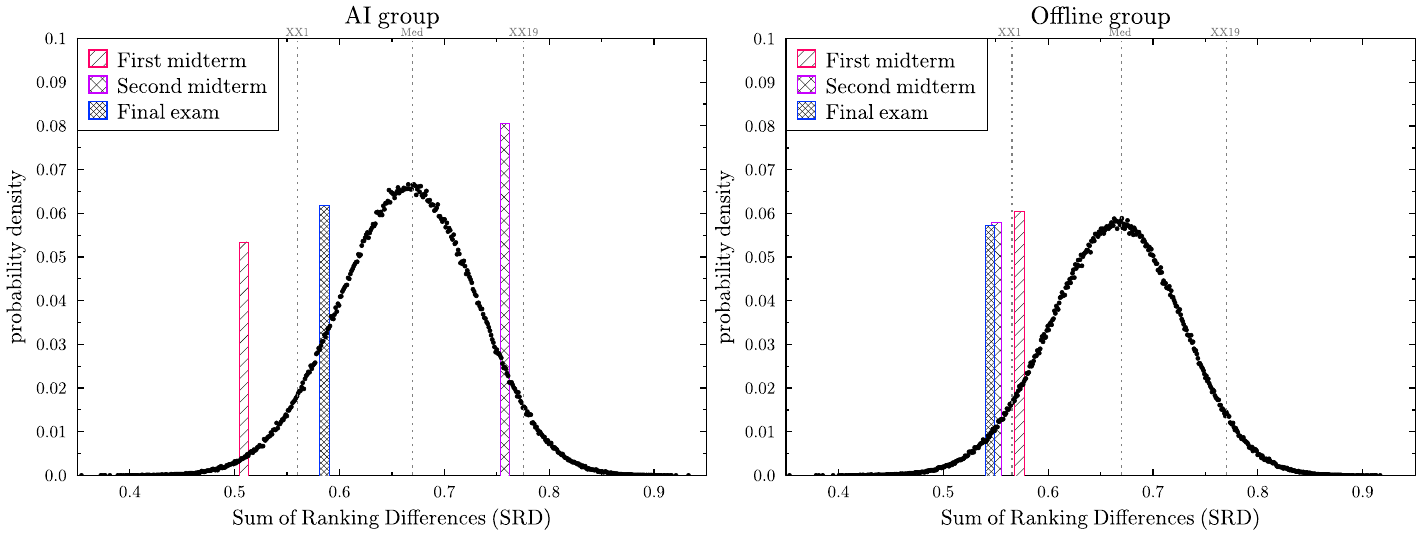}
    \caption{Ranking distance between the paper test and other test scores compared to distances of random rankings measured in SRD. XX1 and XX19 denotes the 0.05 and 0.95 significance threshold respectively. }
    \label{fig:srd}
\end{figure}

Even if the average knowledge level is low, we expect some student to achieve better scores. Some of these performances are due to luck, but certainly there are other cases where motivation or diligence played a role. Thus, if we rank the students according to how many true-or-false questions they guessed correctly, we are close to their true ranking in terms of knowledge. We can measure how this ranking corresponds to rankings derived from other tests. Figure~\ref{fig:srd} shows the Sum of Ranking Difference (SRD) test, which compares the rankings distance between the paper test and other test scores to that of random rankings \citep{Heberger2010,sziklaiApportionmentDistrictingSum2020}.

In case of the AI group, only the first midterm falls to the left of the 0.05 significance threshold (XX1). The other two measurements are indistinguishable from random rankings. In particular the second midterm results are close to the 0.95 threshold (XX19), meaning it ranks the students in almost the exact opposite way. In other words, there is no relation between the performance achieved on the paper test (true knowledge) and the scores obtained at the second midterm and the final exam. For the offline group the SRD values cluster on the left side of the distribution, meaning the students who achieved good scores on the true-or-false test are the one who excelled on other tests as well.

In particular, the final test score for the AI group falls right the 0.05 threshold, while that of the offline group fall left. Hence, Hypothesis~\ref{hypo2} is supported by our data, despite the weakened treatment effect.

\begin{table}[!t]
  \centering
  \begin{threeparttable}
    \caption{Linear regression results}
    \label{tab:lin_reg}
    \begin{tabular}{l *{2}{S[table-format=1.3,table-space-text-post=***]}}
      \toprule
      & \multicolumn{2}{c}{Dependent variable: True-or-False test scores} \\
      \cmidrule(lr){2-3}
      & \multicolumn{1}{c}{AI group} & \multicolumn{1}{c}{Offline group}  \\
      \midrule
      Intercept          & 25.850\sym{***}  & 18.291\sym{***}  \\
                         & 4.471            & 4.150               \\
      First midterm      & 0.573\sym{*}     & 0.009   \\
                         & 0.227            & 0.196                 \\
      Second midterm     & -0.157           & 0.639\sym{**} \\
                         & 0.209            & 0.207                 \\
      Final exam         & -0.031           & 0.132                 \\
                         &  0.086           & 0.095                       \\
      \midrule
      Observations       &  {43}            &  {47}                   \\
      $R^{2}$            &  0.145           &  0.253                    \\
      \bottomrule
    \end{tabular}
    \begin{tablenotes}[flushleft]
      \footnotesize
      \item \sym{*} $p<0.05$, \sym{**} $p<0.01$, \sym{***} $p<0.001$.
    \end{tablenotes}
  \end{threeparttable}
\end{table}

\begin{figure}[!h]
    \centering
    \includegraphics[width=1\linewidth]{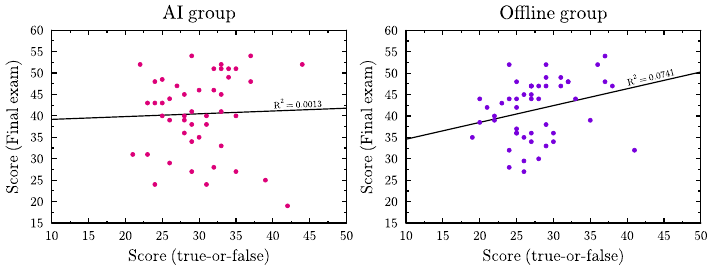}
    \caption{Final exam scores by true-or-false results. $R^2$ values indicate almost no relationship, although the offline group shows a slight visual trend.}
    \label{fig:final_exam_scatter}
\end{figure}

Linear regression results corroborate this finding (see Table~\ref{tab:lin_reg}). The final exam and midterm results explain very little of the variance in the true-or-false test scores, although the offline group is somewhat better in this regard (see also Figure~\ref{fig:final_exam_scatter}). If knowledge were the driving force behind the test scores we would expect to see much higher $R^2$ values. Variance inflation factors (VIF) are close to 1, indicating that the lack of explanatory power is not due to multicollinearity.

In conclusion, although the treatment effect is confounded by the last-minute revision of the experiment, both the SRD test and the linear regression model indicate that the slightly higher true-or-false scores for the AI group are due to random fluctuation, while the the offline group shows genuine learning and a higher---albeit still small---level of accumulated knowledge.

\begin{figure}[!h]
    \centering
    \includegraphics[width=0.6\linewidth]{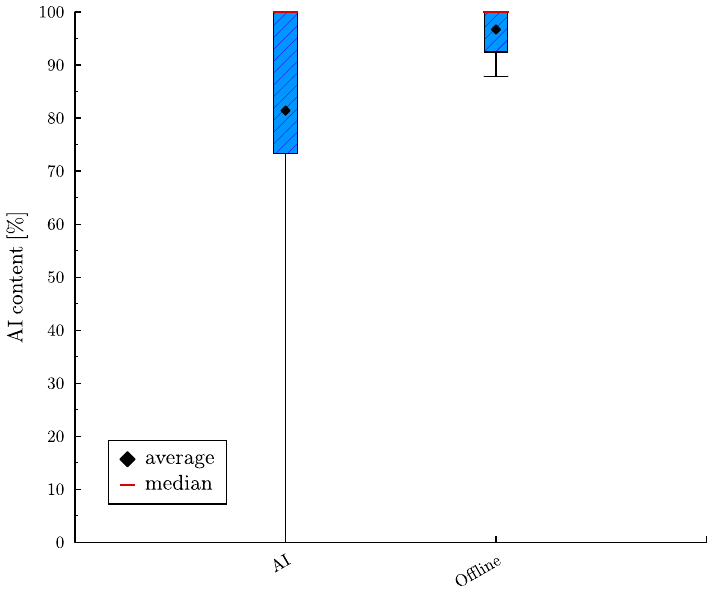}
    \caption{Final exam AI content assessed by an AI detection tool}
    \label{fig:AI_detection}
\end{figure}

To assess the extent to which students relied on AI in the final exam, the authors took a 20\% sample (10 tests from each group) and uploaded into the AI detection tool of CopyLeaks. Among many available AI detection tools studies, such as \cite{Chaka2023} or \cite{Kar2025} for example, suggest that CopyLeaks is one of the most efficient. Figure~\ref{fig:AI_detection} highlights two key findings. First, over-reliance on AI tools is exorbitant, with a median of 100\% in both group. Second, the level of AI usage differed between groups: somewhat surprisingly, the AI group showed slightly lower values. Even when excluding the extreme outlier (one student in the AI group whose test showed no AI usage at all), the average reliance remained lower. Small fluctuations in AI usage can be due to the low sample size, but can also happen due to psychological factors. Student in the offline group who were prohibited using AI assistance in the first two test, were more eager to exploit these tools.

\section{Conclusion}

There is a significant fracture line in the literature regarding the impact of AI tools on students' learning processes. Section~\ref{sec:controversy} lists several factors that may contribute to the discrepancy in experimental results, among which incentivization appears to be the most likely culprit. Experiments with natural incentives, such as students' actual grades, are inherently more difficult to conduct, but paint a more realistic picture on the impact of AI tools.

In Section~\ref{sec:exp_lessons}, we have already summarized the lessons learned about the experimental setting and how to mitigate students' extreme reactions. Perception, group dynamics, and the existence of outside options seem to be crucial elements. Alternative designs, in which students are not divided into fixed roles---or periodically switch roles during the experiment---could help prevent negative reception.

The experimental data lend themselves to some strong conclusions; the following points summarize our key findings.

\begin{itemize}
    \item Uncontrolled use of AI leads to a substantial decrease in knowledge. The data indicate an effect likely between 20 and 40 \textit{percentage points} compared to previous years. It is not clear how AI usage affect learning in more controlled settings (e.g.\ tutoring), when students have limitations in how they can interact with AI.
    \item Open book computer lab setting seems to be the among the most exposed examination formats, although psychological models prognosticate that the whole learning process might be affected regardless of whether AI assistance is available on the exams or not.
    \item The reliance on AI tools and attitude toward responsible usage seems to also depend on the discipline. Difficult subjects---such as mathematics, which student often perceive as having little practical value---are especially vulnerable to the misuse of AI.
    \item Despite claims of heavy competition for the stipends, very few students put effort into engaging with the material. Examination of the test sheets indicates that the large majority of students were content to cede control and rely excessively on AI tools.
    \item Over-reliance is also reflected in the extreme reactions to the experiment. Even only after a few years of usage, students cannot fathom mastering a difficult subject without the assistance of AI tools.
\end{itemize}

These findings warrant a more scrutinized look on the impact of AI tools on learning outcomes and caution against overly optimistic approaches that seek to integrate this innovation into education prematurely.

\pagebreak
\section*{Appendix}

\textbf{Paper test question example 1}

\begin{table}[!h]
\begin{tabular}{llll}
\multirow{2}{*}{} & Which of the following are true about the solution sets                                   & \multirow{2}{*}{True} & \multirow{2}{*}{False} \\
                  & of linear programming problems: An LP …                                                   &                       &                        \\ \cline{3-4}
a)                & \multicolumn{1}{l|}{may have no feasible solution.}                                       & \multicolumn{1}{l|}{~} & \multicolumn{1}{l|}{~}  \\ \cline{3-4}
b)                & \multicolumn{1}{l|}{if it has a feasible solution, then it also has an optimal solution.} & \multicolumn{1}{l|}{~} & \multicolumn{1}{l|}{~}  \\ \cline{3-4}
c)                & \multicolumn{1}{l|}{may have only one optimal solution.}                                  & \multicolumn{1}{l|}{~} & \multicolumn{1}{l|}{~}  \\ \cline{3-4}
d)                & \multicolumn{1}{l|}{may have a bounded objective function but no optimal solution.}       & \multicolumn{1}{l|}{~} & \multicolumn{1}{l|}{~}  \\ \cline{3-4}
e)                & \multicolumn{1}{l|}{may have exactly two corner-point (vertex) optimal solutions.}        & \multicolumn{1}{l|}{~} & \multicolumn{1}{l|}{~} \\ \cline{3-4}
\end{tabular}
\end{table}

\textbf{Paper test question example 2}

\begin{figure}[!h]
\begin{minipage}{0.55\linewidth}
    \centering
    \includegraphics[width=0.7\linewidth]{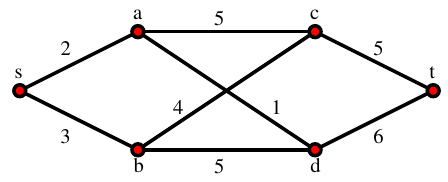}
    \caption{Example graph. Edge weights indicate construction costs.}
    \label{fig:paper_test_example}
\end{minipage}
\begin{minipage}{0.4\linewidth}
\begin{tabular}{llll}
   &                                       & \textbf{True}                  & \textbf{False} \\ \cline{3-4}
a) & \multicolumn{1}{l|}{9}                & \multicolumn{1}{l|}{~} & \multicolumn{1}{l|}{~} \\ \cline{3-4}
b) & \multicolumn{1}{l|}{not more than 13} & \multicolumn{1}{l|}{~} & \multicolumn{1}{l|}{~} \\ \cline{3-4}
c) & \multicolumn{1}{l|}{16}               & \multicolumn{1}{l|}{~} & \multicolumn{1}{l|}{~} \\ \cline{3-4}
d) & \multicolumn{1}{l|}{at least 15}      & \multicolumn{1}{l|}{~} & \multicolumn{1}{l|}{~} \\ \cline{3-4}
e) & \multicolumn{1}{l|}{not more than 20} & \multicolumn{1}{l|}{~} & \multicolumn{1}{l|}{~} \\ \cline{3-4}
\end{tabular}
\end{minipage}
\end{figure}

The cost of the minimum cost spanning tree of the graph in Figure~\ref{fig:paper_test_example} is

\bibliography{references}

\begin{thebibliography}{}

\bibitem[Al-Zahrani, 2024]{Al-Zahrani2024}
Al-Zahrani, A.~M. (2024).
\newblock Unveiling the shadows: Beyond the hype of {AI} in education.
\newblock {\em Heliyon}, 10(9).

\bibitem[{AsPredicted}, 2025]{aspredicted223975}
{AsPredicted} (2025).
\newblock Impact of {AI} tools on learning outcomes, {B}udapest, {J}une 2025.
\newblock \url{https://aspredicted.org/z68g-wqzc.pdf}.
\newblock AsPredicted \#223975.

\bibitem[Biwer et~al., 2020]{Biwer2020}
Biwer, F., de~Bruin, A. B.~H., Schreurs, S., and oude Egbrink, M. G.~A. (2020).
\newblock Future steps in teaching desirably difficult learning strategies:
  Reflections from the study smart program.
\newblock {\em Journal of Applied Research in Memory and Cognition},
  9(4):439--446.

\bibitem[Booij et~al., 2016]{Booij2016}
Booij, A.~S., Leuven, E., and Oosterbeek, H. (2016).
\newblock Ability peer effects in university: Evidence from a randomized
  experiment.
\newblock {\em The Review of Economic Studies}, 84(2):547--578.

\bibitem[Chaka, 2023]{Chaka2023}
Chaka, C. (2023).
\newblock Detecting ai content in responses generated by {ChatGPT, YouChat, and
  Chatsonic}: The case of five {AI} content detection tools.
\newblock {\em Journal of Applied Learning and Teaching}, 6(2):94--104.

\bibitem[de~Bruin et~al., 2023]{deBruin2023}
de~Bruin, A. B.~H., Biwer, F., Hui, L., Onan, E., David, L., and Wiradhany, W.
  (2023).
\newblock Worth the effort: the start and stick to desirable difficulties
  (s2d2) framework.
\newblock {\em Educational Psychology Review}, 35(2):41.

\bibitem[Favero et~al., 2025]{favero2025}
Favero, L., Pérez-Ortiz, J.-A., Käser, T., and Oliver, N. (2025).
\newblock Do {AI} tutors empower or enslave learners? toward a critical use of
  {AI} in education.

\bibitem[Georgiou, 2025]{Georgiou2025}
Georgiou, G.~P. (2025).
\newblock {ChatGPT} produces more "lazy" thinkers: Evidence of cognitive
  engagement decline.

\bibitem[Gerlich, 2025]{Gerlich2025}
Gerlich, M. (2025).
\newblock {AI} tools in society: Impacts on cognitive offloading and the future
  of critical thinking.
\newblock {\em Societies}, 15(1).

\bibitem[Hal{\'a}sz, 2025]{halasz2025_telex}
Hal{\'a}sz, N. (2025).
\newblock A hallgatók egy része használhat {MI}-t a vizsgán, a másik
  része nem, fel is háborodtak.
\newblock {\em Telex}.
\newblock [In Hungarian] Accessed: 2025-10-15
  \url{https://telex.hu/belfold/2025/05/22/corvinus-egyetem-hallgatok-oktatok-mesterseges-intelligencia-kiserlet}.

\bibitem[Harasym et~al., 2008]{Harasym2008}
Harasym, P.~H., Woloschuk, W., and Cunning, L. (2008).
\newblock Undesired variance due to examiner stringency/leniency effect in
  communication skill scores assessed in osces.
\newblock {\em Advances in Health Sciences Education}, 13(5):617--632.

\bibitem[H{\'e}berger, 2010]{Heberger2010}
H{\'e}berger, K. (2010).
\newblock Sum of ranking differences compares methods or models fairly.
\newblock {\em TrAC Trends in Analytical Chemistry}, 29(1):101--109.

\bibitem[Kar et~al., 2025]{Kar2025}
Kar, S.~K., Bansal, T., Modi, S., and Singh, A. (2025).
\newblock How sensitive are the free ai-detector tools in detecting
  ai-generated texts? a comparison of popular ai-detector tools.
\newblock {\em Indian journal of psychological medicine}, 47(3):275--278.

\bibitem[Karpicke et~al., 2009]{Karpicke2009}
Karpicke, J.~D., Butler, A.~C., and III, H. L.~R. (2009).
\newblock Metacognitive strategies in student learning: Do students practise
  retrieval when they study on their own?
\newblock {\em Memory}, 17(4):471--479.

\bibitem[Kestin et~al., 2025]{Kestin2025}
Kestin, G., Miller, K., Klales, A., Milbourne, T., and Ponti, G. (2025).
\newblock Ai tutoring outperforms in-class active learning: an rct introducing
  a novel research-based design in an authentic educational setting.
\newblock {\em Scientific Reports}, 15(1):17458.

\bibitem[Kolen and Brennan, 2014]{kolen2014test}
Kolen, M.~J. and Brennan, R.~L. (2014).
\newblock {\em Test Equating, Scaling, and Linking: Methods and Practices}.
\newblock Springer Science+Business Media, New York, NY, 3rd edition.

\bibitem[Kool et~al., 2010]{kool2010}
Kool, W., McGuire, J.~T., Rosen, Z.~B., and Botvinick, M.~M. (2010).
\newblock Decision making and the avoidance of cognitive demand.
\newblock {\em Journal of Experimental Psychology: General}, 139(4):665--682.

\bibitem[Kosmyna et~al., 2025]{kosmyna2025}
Kosmyna, N., Hauptmann, E., Yuan, Y.~T., Situ, J., Liao, X.-H., Beresnitzky,
  A.~V., Braunstein, I., and Maes, P. (2025).
\newblock Your brain on chatgpt: Accumulation of cognitive debt when using an
  ai assistant for essay writing task.

\bibitem[Rodrigues, 2025]{Rodrigues2025}
Rodrigues, D. (2025).
\newblock The ai revolution in education: From cognitive enhancement to hybrid
  intelligence.
\newblock \url{https://ssrn.com/abstract=5327624}.
\newblock Available at SSRN.

\bibitem[Sziklai and H{\'e}berger,
  2020]{sziklaiApportionmentDistrictingSum2020}
Sziklai, B.~R. and H{\'e}berger, K. (2020).
\newblock Apportionment and districting by {{Sum}} of {{Ranking Differences}}.
\newblock {\em PLOS ONE}, 15(3):e0229209.

\bibitem[Wang and Fan, 2025]{Wang2025}
Wang, J. and Fan, W. (2025).
\newblock The effect of chatgpt on students' learning performance, learning
  perception, and higher-order thinking: insights from a meta-analysis.
\newblock {\em Humanities and Social Sciences Communications}, 12(1):621.

\bibitem[Westbrook et~al., 2013]{Westbrook2013}
Westbrook, A., Kester, D., and Braver, T.~S. (2013).
\newblock What is the subjective cost of cognitive effort? load, trait, and
  aging effects revealed by economic preference.
\newblock {\em PLOS ONE}, 8(7):1--8.

\bibitem[Zrubka et~al., 2025]{Zrubka2025ModiBot}
Zrubka, Z., Marczis{\'a}k, B., S{\"u}li, P., Eigner, G., Benedek, M.,
  G{\'a}sp{\'a}r, J., and Sziklai, B. (2025).
\newblock Piloting {ModiBot}: A large language model-based moderator in normal
  and emotionally challenging focus group interactions.
\newblock In {\em Proceedings of the 25th IEEE International Symposium on
  Computational Intelligence and Informatics (CINTI 2025)}, pages 1--8,
  Budapest, Hungary. \'Obuda University, IEEE.

\end{thebibliography}


\end{document}